\documentclass[%
 aip,
 amsmath,amssymb,
 reprint,%
]{revtex4-1}

\usepackage{graphicx}
\usepackage{dcolumn}
\usepackage{bm}

\usepackage[utf8]{inputenc}
\usepackage[T1]{fontenc}
\usepackage{mathptmx}
\usepackage{etoolbox}
\usepackage{xcolor}
\DeclareMathAlphabet{\mathcal}{OMS}{cmsy}{m}{n}

\makeatletter
\def\@email#1#2{%
 \endgroup
 \patchcmd{\titleblock@produce}
  {\frontmatter@RRAPformat}
  {\frontmatter@RRAPformat{\produce@RRAP{*#1\href{mailto:#2}{#2}}}\frontmatter@RRAPformat}
  {}{}
}%
\makeatother
\begin{document}

\preprint{AIP/123-QED}

\title[]{Benchmarking mixed quantum-classical dynamics for collective electronic strong coupling}
\author{Arun Kumar Kanakati}
\email{arun.k.kanakati@jyu.fi}
\affiliation{
  Nanoscience Center and Department of Chemistry, University of
  Jyv\"{a}skyl\"{a}, P.O. Box 35, 40014 Jyv\"{a}skyl\"{a},
  Finland.}

\author{Oriol Vendrell}
\affiliation{
    Theoretical Chemistry, Institute of Physical Chemistry, Heidelberg University, Im Neuenheimer Feld 229, 69120 Heidelberg, Germany.}

\author{Gerrit Groenhof}%
 \email{gerrit.x.groenhof@jyu.fi}
\affiliation{
  Nanoscience Center and Department of Chemistry, University of
  Jyv\"{a}skyl\"{a}, P.O. Box 35, 40014 Jyv\"{a}skyl\"{a},
  Finland.}
%

\date{\today}

\begin{abstract}
Experiments indicate that collective coupling of molecular ensembles to confined optical modes can modify excited-state dynamics and photochemical reactivity. To describe such cavity-induced effects at atomic resolution, semi-classical molecular dynamics approaches have been developed that treat nuclear motion classically while describing the collective light-matter interaction within the Tavis-Cummings framework of quantum electrodynamics. Here, we benchmark mixed quantum-classical approaches, Ehrenfest dynamics and Fewest-Switches Surface Hopping (FSSH), for simulating nonadiabatic dynamics of electronically strongly coupled carbon monoxide molecules. Their predictions are compared against numerically exact quantum dynamics simulations performed with the multi-configuration time-dependent Hartree (MCTDH) method, which treats both electronic and nuclear degrees of freedom quantum mechanically. We find that the semi-classical approaches reproduce the qualitative features of the full quantum dynamics. Quantitative agreement is best achieved with FSSH when a decoherence correction is included. These results demonstrate that mixed quantum-classical methods provide a computationally efficient and quantitatively reliable alternative to fully quantum simulations for investigating nonadiabatic photochemistry under collective electronic strong coupling in systems beyond the reach of exact quantum treatments.
\end{abstract}

\maketitle

%

\section{\label{section:intro}Introduction}


Over the past decades, significant efforts have focused on engineering materials capable of precisely controlling the properties of light.\cite{del2019light,malinauskas2016ultrafast,perez2025large} Conversely, controlling the intrinsic properties of materials using confined light fields presents both a major challenge and a profound opportunity. Recent experiments suggest that optical resonators, such as cavities and plasmonic lattices, may provide such control. Indeed, embedding materials inside optical cavities has been demonstrated to reshape their physico-chemical properties, influencing energy transport,\cite{coles2014polariton,Lerario2017high,Rozenman2018long,Georgiou2021ultralong,Berghuis2022controlling,Forrest2020ultralong,RajPandya2022tuning,Balasubrahmaniyam2023enhanced,IliaJPCL2025,krupp2025quantum} charge mobility,\cite{Orgiu2015conductivity,Krainova2020polaron,nagarajan2020conductivity,Bhatt2021enhanced} lasing thresholds,\cite{Kena-Cohen2010room,Hakala2018bose} and even photo-chemistry.\cite{hutchison2012modifying,Munkhbat2018,Stranius2018selective,Mony2021,Yu2021barrier} The possibility of harnessing cavity effects to steer photochemical reactions could open the door to transformative applications in artificial light harvesting, energy storage, and quantum technologies.\cite{Bhuyan2023} Yet, despite its promise, progress in this new field of polaritonic chemistry\cite{ribeiro2018polariton} is significantly hampered  by a lack of theoretical understanding.  

Optical resonators, such as a Fabry-P\'{e}rot microcavity, enhance light-matter interactions by confining the electromagnetic field into very small volumes.\cite{Vahala2003} If the strength of the light-matter interaction becomes sufficiently high, which can be achieved by increasing confinement or by increasing the \emph{collective} oscillator strength of the material with more molecules, the system can enter the \emph{strong coupling} regime, where material transitions (electronic or vibrational) hybridize with the confined photon modes of the optical resonator.\cite{Torma2015strong,Garcia-Vidal2021} These light-matter hybrid states are called \emph{polaritons} and are characterized by a Rabi splitting of the coupled material's absorption spectrum into an upper and lower polariton. Because these polaritonic states are coherent superpositions of molecular transitions and cavity modes, an excitation is delocalized over many molecules. In addition to these bright and delocalized polaritonic states, also "dark" states form, which are the remaining superpositions that lack cavity mode contribution.

Because changes in material properties have so far only been observed in combination with Rabi splitting, these changes have been attributed to polaritons.\cite{Garcia-Vidal2021} However, there is currently no consensus on the microscopic mechanisms by which polariton formation could influence photochemical reactivity, in particular, because the large majority of states are dark and hence  similar to the uncoupled molecular states.\cite{scholes2020entropy,Dutta2024}

A central conceptual challenge is that, under collective strong coupling, an excitation is coherently delocalized over many molecules, whereas chemical reactivity is intrinsically local.\cite{ribeiro2018polariton} Resolving this apparent contradiction requires theoretical approaches that can simultaneously describe collective light-matter coherence and local nonadiabatic nuclear dynamics with chemical accuracy. To circumvent modeling large numbers of molecules, most theoretical approaches focus on single molecules instead, using much stronger cavity vacuum fields.\cite{Kowalewski2016a,Flick2017cavity,Fregoni2018,Haugland2020CCT,Fabri2021born,Schaefer2022,li2021collective,li2021cavity,Sun2022suppression,Li2022,Lee2025,Bauman2025}. Because the Rabi splitting scales with the number of molecules, $N_\text{mol}$, as $\sqrt{N_\text{mol}}$,\cite{Torma2015strong} these fields are enhanced by the same factor to achieve strong coupling with only a single molecule. However, such scaled fields can easily become nonphysical,\cite{delapradilla2025} and hence induce changes to the photo-chemistry that are not real.\cite{krupp2026PRR} On the other hand, models from quantum optics that focus on describing collective strong coupling in large $N$ limit,\cite{herrera2016cavity,Ahn2023,Perez-Sanchez2023} tend to lack chemical details that are needed to unravel how the coherent coupling impacts the chemistry locally.

To overcome the limitations of these two modeling extremes for modeling electronic strong coupling, a divide-and-conquer strategy was proposed.\cite{galego2015cavity,Kowalewski2016b,luk2017multiscale} Combining the Born-Oppenheimer and long-wavelength approximations, polaritonic states are obtained within the Tavis-Cummings (TC) framework of quantum optics,\cite{Tavis_PhysRev_1969} using the adiabatic electronic ground and excited states of the molecules evaluated at a suitable level of quantum chemistry, as the basis. 

By propagating the nuclear degrees of freedom classically under the influence of the polaritonic wavefunction, while simultaneously propagating that wavefunction as a time-dependent superposition of the TC eigenstates within a mixed quantum-classical framework, the molecular dynamics (MD) in the collective strong coupling regime can be modeled with chemical accuracy.\cite{Sokolovskii2024b} Through extensive parallelization, such semi-classical MD simulations of thousands of molecules in Fabry-P\'{e}rot micro-cavities helped resolve important questions, such as: (\textit{i}) why, despite the short lifetimes of cavity modes, the polariton appears long-lived;\cite{Groenhof2019,tichauer2021multi} (\textit{ii}) why polariton transport is not ballistic but diffusive\cite{sokolovskii2023multi,IliaJPCL2025} and can even reverse on longer timescales;\cite{tichauer2023tuning} and (\textit{iii}) what is the role of molecular disorder on polaritonic effects.\cite{Dutta2024}

However, because these approaches propagate the nuclear degrees of freedom according to classical Newtonian dynamics, they neglect nuclear quantum effects and quantum coherence in the vibrational motion. While such mixed quantum-classical schemes have proven successful for modeling large polaritonic systems, their quantitative accuracy for nonadiabatic dynamics under collective electronic strong coupling has not been systematically assessed. 

Here, we therefore benchmark semi-classical polaritonic dynamics against numerically exact quantum simulations. To render the latter tractable, we consider up to five carbon monoxide (CO) molecules collectively coupled to a single-mode optical cavity resonant with the electronic transition to the $^1\Pi$ excited state. By comparing observables obtained from Ehrenfest and Fewest-Switches Surface Hopping dynamics\cite{Tully1990,Tully1991} to those from multi-configuration time-dependent Hartree (MCTDH) simulations,\cite{Manthe_jcp19923199} we quantify the impact of the classical nuclear approximation on polaritonic nonadiabatic dynamics.

The paper is organized as follows: In section~\ref{section:sim_details}, we present our model of CO molecules in a single-mode optical cavity and share the details of our simulations on this system. Then, in section~\ref{section:results} we compare the observables obtained from semi-classical and quantum dynamics simulations for a varying number of CO molecules coupled to the cavity. Finally, we conclude in section~\ref{sec:conclusions} with a summary and outlook.

\section{Simulation details}\label{section:sim_details}

\subsection{Potential energy surfaces}\label{sec:abintio_calc}

The equilibrium geometry of the electronic ground state (S$_0$) of CO was optimized with the Gaussian09 program\cite{G09} at the coupled-cluster singles and doubles (CCSD) level of \textit{ab initio} theory employing the augmented correlation-consistent polarized valence double zeta (aug-cc-pVDZ) basis set. The S$_0$ minimum energy geometry has a $C_{\infty v}$ point group symmetry with an equilibrium bond length of 1.1405~{\AA} and a fundamental anharmonic frequency of 2172~$\text{cm}^{-1}$.

\begin{figure*}[!t]
\centering
\includegraphics[width=1\textwidth ]{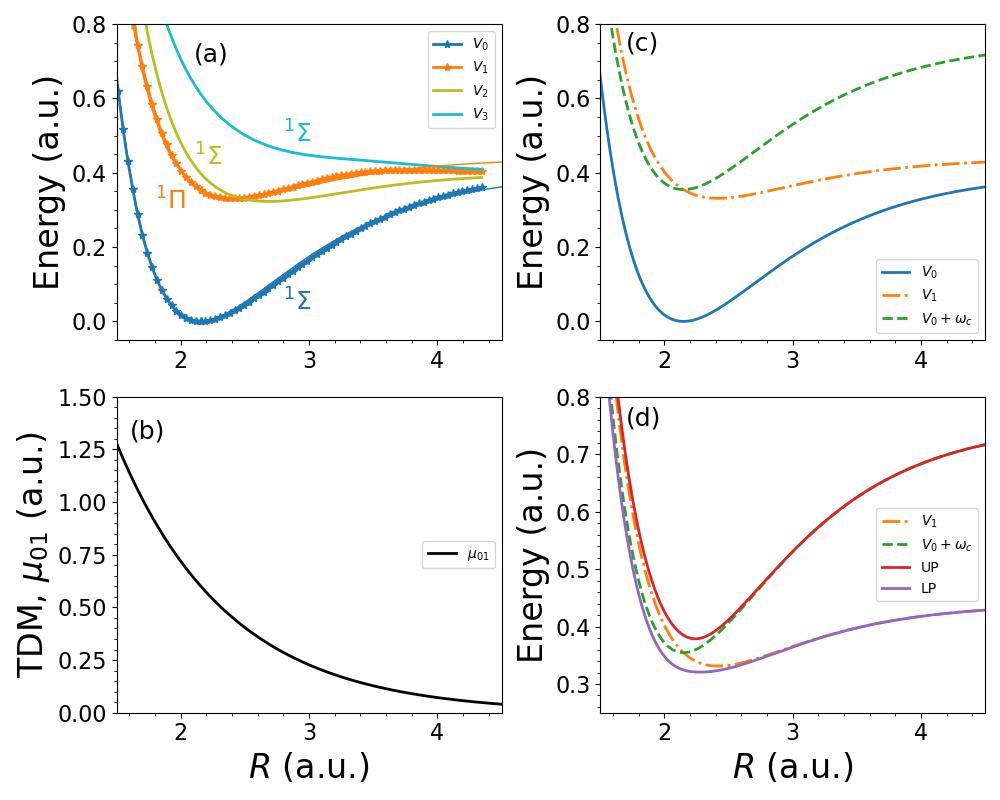}
\caption{\label{fig:CO_pPESs}Adiabatic potential energy profiles of the bare CO molecule (a), and transition dipole moment between the ground, $V_0$ ($^1\Sigma$) and first excited, $V_1$ ($^1\Pi$), electronic states as a function of inter-nuclear distance, $R$ (b). Potential energy profile of one CO molecule in a cavity: The coupling strength is equal to zero (c) and 0.050 au (d). The profile without coupling shows the energy of the ground state shifted by the cavity mode energy (\textit{i.e.}, cavity mode excitation, $V_0+\omega_c$), as well as the ground and excited electronic states of the CO molecule without cavity mode excitation. Panel (d) shows the potential energy profiles of hybrid states that form when the coupling between the cavity and the molecule is 0.050 au. These profiles were obtained by diagonalizing the Tavis-Cummings Hamiltonian [cf. Eq. \ref{eq:arrow_matrix}].}
\end{figure*}

The potential energy curves and dipole moments as a function of the CO bond length, $R$, were calculated with the MOLPRO program package\cite{Molpro2010} at the CASSCF(14,10)/aug-cc-pVDZ level of multi-configuration self-consistent field theory, with state-averaging over the six lowest-energy singlet electronic states. Energies, dipole moments, and transition dipole moments (TDM) were calculated for eighty internuclear distances between 0.7 \AA ~($\sim$ 1.3 a.u.) and 2.3 \AA ~($\sim$ 4.3 a.u.). The energy profiles for electronic ground state ($^1\Sigma$) and for one of the doubly degenerate electronic excited states ($^1\Pi$) are shown in Fig.~\ref{fig:CO_pPESs}a, along with the transition dipole moment between these states [cf. Fig.~\ref{fig:CO_pPESs}b]. 

A Morse potential was fitted to each potential energy profile [cf. Fig. \ref{fig:CO_pPESs}(a)]:
\begin{equation}
  V_i(R) =  E_i + D_i\left[1-\exp\left(-\alpha_i(R-R^\text{eq}_i)\right)\right]^2\label{eq:Morse}
\end{equation} 
with $R^\text{eq}_i$ is the equilibrium C-O distance, and $D_i$ the dissociation energy in the ground ($i=0$) and excited state ($i=1$), respectively. The values of these fitting parameters are provided in Table \ref{tab_fit_params}. The same fitted Morse potentials and transition dipole curves were used in both the MCTDH and semi-classical simulations to ensure that differences arise solely from the treatment of nuclear dynamics.

\begin{table}[h!] 
\centering
\caption{\label{tab_fit_params}Morse potential fitting parameters (in a.u.) for both ground $^1\Sigma^+$ and first excited $^1\Pi$ electronic states.} 
\begin{tabular}{|c|c|c|c|c|c|}
\hline  \hline
 Parameters &  $^1\Sigma$ & $^1\Pi$     \\
\hline 
$E_i$ & ~0.0000 & ~0.3311  \\
$D_i$ & ~0.4013 & ~0.1086  \\
$\alpha_i$ & 1.2710 & 1.4144 \\
$R^\text{eq}_i$ &  2.15   &  2.42   \\
\hline \hline
\end{tabular}
\end{table} 

\subsection{Quantum dynamics simulations of strongly coupled CO molecules} \label{quantum}

\subsubsection{Molecule-cavity Hamiltonian}

We consider a one-dimensional array of $N_\text{mol}$ identical and non-interacting CO molecules inside an optical cavity. 
The molecular ensemble-cavity Hamiltonian contains a molecular part,  cavity part and their interaction:
\begin{equation} \label{eq:Ham_total}
    \hat{H} = \sum_{k=1}^{N_\text{mol}} \hat{H}_\text{mol}^{(k)} + \hat{H}_\text{cav} + \hat{H}_\text{cav-mol}    
\end{equation}
with  $\hat{H}_\text{mol}^{(k)} = \hat{T}_n^{(k)} + \hat{H}_e^{(k)}$ the Hamiltonian of the $k^\text{th}$ CO molecule, including the kinetic and potential energy operators of the bare molecule:
\begin{eqnarray}
    \hat{H}_\text{mol} = -\frac{\hbar^2}{2\mu}\frac{\partial^2}{\partial R^2}{\bf{1}} + 
    \begin{pmatrix}
    V_0(R)  &  0 \\ 
    0       & V_1(R)
    \end{pmatrix}
\end{eqnarray}
where $\mu$ is the reduced nuclear mass, $R$ the C-O internuclear distance, and  $V_0(R)$ and $V_1(R)$ the potential energy profiles in the electronic ground and excited states, respectively (Fig.~\ref{fig:CO_pPESs}a), fitted to Morse potentials (Equation~\ref{eq:Morse}). 

Within the long-wavelength approximation, the interaction between a molecule and the cavity mode is described with the Rabi Hamiltonian (\textit{i.e.}, the Pauli-Fierz Hamiltonian \emph{without} dipole-self energy):\cite{faisal1987theory,galego2015cavity,Kowalewski2016a,Flick2017atoms,Flick2017cavity,vendrell2018coherent,vendrell2018collective}
\begin{equation}
    \hat{H}_\text{cav} + \hat{H}_\text{cav-mol} = \hbar\omega_\text{c}\left(\frac{1}{2}+\hat{a}^{\dagger}\hat{a} \right) + g\overrightarrow{\hat{D}}\cdot\overrightarrow{\epsilon}_c\left(\hat{a}^{\dagger}+\hat{a} \right)\label{eq:cavityhamiltonian}
\end{equation}
Here, $\omega_c$ the cavity mode frequency, $\hat{a}^\dagger$ and $\hat{a}$ the photon creation and annihilation operators, respectively; $g$ the coupling strength defined as $g= \sqrt{\frac{\hbar\omega_c}{2V\epsilon_0}}$;  $\overrightarrow{\hat{D}}$ the molecular dipole operator, and $\overrightarrow{\epsilon}_c$ the cavity mode polarization. For simplicity, we only consider a single polarization direction.

\subsubsection{Quantum dynamics propagation}

In the quantum dynamics simulations, the time evolution of the molecular ensemble-cavity wave function is computed using the multi-configuration time-dependent Hartree (MCTDH) approach,\cite{Meyer_cpl199073,BECK3242000} implemented in the Heidelberg MCTDH program version 8.6.2.\cite{mctdh_pack85} In this approach, the wave function is approximated as a product of time-dependent coefficients with time-dependent basis functions for each nuclear, electronic, and cavity degree of freedom. For convenience, the cavity mode is treated as an harmonic oscillator in terms of the position ($\hat{q}_\text{c}=\sqrt{\hbar/2\omega_c}[\hat{a}^\dagger + \hat{a}]$) and momentum operators ($\hat{p}_\text{c}= i\sqrt{\hbar\omega_\text{c}/2}[\hat{a}^\dagger - \hat{a}]$) rather than the annihilation and creation operators in Equation~\ref{eq:cavityhamiltonian}.\cite{vendrell2018coherent,vendrell2018collective,ulusoy2019modifying,ulusoy2020dynamics} The method was first introduced to treat the multi-dimensional quantum dynamics of molecular systems.\cite{Meyer_cpl199073,manthe1992multiconfigurational,Manthe_jcp19923199} An in depth review of the basic theory can be found in Beck \textit{et al.}\cite{BECK3242000}   

The basic theory is briefly described in this section, where the usual nomenclature in the MCTDH literature is used for consistency.\cite{BECK3242000} The MCTDH ansatz for the wave function reads
\begin{eqnarray}
    \vert\Psi(Q_1,...,Q_f,t)\rangle &=& \sum_{j_1}^{n_1}...\sum_{j_f}^{n_f}A_{j_{1}...j_{f}}(t) \prod_{\kappa=1}^f \vert\phi_{j_\kappa}^{(\kappa)}(q_{\kappa},t)\rangle \\
                                &=&  \sum_J A_J(t)\vert\Phi_J(t)\rangle,     
\end{eqnarray}
where $A_J(t)$ is the time-dependent expansion coefficient of the \textit{J}-th configuration labeled with multi-index \textit{J}, and $\vert\Phi_J(t)\rangle $ is the \textit{J}-th time-dependent Hartree product, which is a direct product of single-particle functions (SPFs) for each degree of freedom. These are analogous to molecular orbitals in electronic structure theory.\cite{szabo2012modern} The SPFs are expanded in turn in a time-independent basis for each degree of freedom
\begin{equation}
    \vert\phi_{j_{\kappa}}^{(\kappa)}(q_{\kappa},t)\rangle = \sum_{i=1}^{N_{\kappa}}c_{i,j}^{(\kappa)}(t)\vert\chi_i^{(\kappa)}(q_{\kappa})\rangle.  
\end{equation}
where, for convenience, very often the states of the primitive representation $|\chi_i^{(\kappa)}\rangle$ are taken from a discrete variable representation.\cite{BECK3242000} 

For molecules coupled to a cavity mode, the MCTDH wave function becomes:
\begin{widetext}
\begin{eqnarray} \label{eq:cav-mol_WF}
    \lvert\Psi(t)\rangle = \sum_{j_1,...,j_N,j_p}^{n_1,...,n_N,n_p}A_{j_1,...,j_N,j_p}\cdot\prod_{l=1}^{N_\text{mol}}\left(\sum_{s_l=1}^{N_s}\phi_{s_l,j_l}^{(l)}(t)\lvert\psi_{s_l}^{(l)}\rangle \right)\left(\sum_{P=1}^{N_p}B_{P,j_p}(t)\lvert P\rangle \right).
\end{eqnarray}
\end{widetext}
Here, $n_l$ and $n_p$ denote the number of SPF basis for each molecule and the cavity mode, respectively, where the nuclear and electronic degrees of freedom are combined into a single logical mode. As previously, $N_s$ represents the number of relevant electronic states per molecule, and $N_p$ specifies the maximum number of photons permitted within the cavity. The $\phi_{s_l,j_l}^{(l)}(t)$ functions represent the nuclear wave packets for molecule $l$ in electronic state $s_l$, with the index $j_l$ indicating the configuration space specified by $A_{j_1,…,j_N,j_p}$ in Equation \ref{eq:cav-mol_WF}. Conversely, $B_{P,j_p(t)}$ denote the expansion coefficients of the primitive basis functions of the $P$ cavity photons with configuration space index $j_p$.  

Using the Dirac–Frankel variational principle,\cite{BECK3242000} equations of motion for the expansion coefficients and SPFs were derived.\cite{Manthe_jcp19923199,Meyer_cpl199073,BECK3242000} These are coupled differential equations for the expansion coefficients and SPFs. The wave packet is optimally described with this and the time-dependent basis moves with the wave packet keeping the basis size small. These are $n_{\kappa}$ SPFs for $\kappa$ DOF which are represented in terms of $N_{\kappa}$ primitive basis functions or grid points. The efficiency of the MCTDH algorithm grows with the ratio $N_{\kappa}/n_{\kappa}$. The method uses discrete variable representation (DVR) in combination with the fast Fourier transform algorithm and powerful integrators for the evolution of the wavepacket on the grid. 

To facilitate comparison with semi-classical molecular dynamics simulations, in which the photo-excitation is modeled as an instantaneous population transfer into one of the polaritonic states, we generate the initial state for the MCTDH simulations through application of the operator $\hat{T}_{\pm}$, which directly excites the cavity-molecule system into a 1:1 light-matter superposition state (with the minus sign in the index for LP and plus for UP):\cite{ulusoy2019modifying} 
\begin{equation}
    \hat{T}_{\pm} = \frac{1}{\sqrt{2}}\left(\hat{a}^{\dagger}+\hat{a} \right)\mp \sum_{\kappa}^{N_\text{mol}} \frac{1}{\sqrt{2N_\text{mol}}}\left(|0_{\kappa}\rangle\langle 1_{\kappa}|+ \text{h.c.}  \right) \label{T-oper}. 
\end{equation}
where $|0_{\kappa}\rangle$ and $|1_{\kappa}\rangle$ are the two electronic states of the $\kappa^\text{th}$ CO molecule coupled to the optical cavity mode. This operator generates the lower (LP) and upper (UP) polariton states corresponding to the equilibrium nuclear geometry.

\subsubsection{Wavefunction analysis}

The populations of the UP and LP states are the key observables characterizing the time-evolution of the hybrid system. A convenient way to extract such information from the MCTDH wavefunction is to compute the expectation value of the light-matter interaction term in the Hamiltonian divided by the interaction strength:\cite{ulusoy2019modifying,ulusoy2020dynamics}
\begin{equation} \label{eq_vint}
    V_\text{int,r} = \langle\Psi(t)|\hat{O}|\Psi(t)\rangle 
\end{equation}
with 
\begin{equation}
    \hat{O} = -\left(\hat{a}^{\dagger}+\hat{a}\right)\sum_{\kappa}\left(|0_{\kappa}\rangle\langle 1_{\kappa}|+ |1_{\kappa}\rangle\langle 0_{\kappa}|  \right)
\end{equation}
Because LP and UP correspond to symmetric and antisymmetric superpositions of molecular and photonic excitations, the expectation value of the interaction operator serves as a signed measure of polaritonic character.

\subsection{Semi-classical molecular dynamics simulations of strongly coupled CO molecules}\label{subsec:semi-classical}

Within a mixed quantum-classical framework based on the Born-Oppenheimer approximation, we separate the electronic plus cavity mode degrees of freedom from the nuclear degrees of freedom.\cite{galego2015cavity} Neglecting the dipole-self energy, which is very small for realistic cavity setups,\cite{delapradilla2025} and adopting the rotating wave approximation (RWA), valid when the cavity mode and molecular excitation are resonant and the coupling strength is significantly smaller than the excitation energy, we can describe the interactions between $N_\text{mol}$ CO molecules and the cavity mode with the Tavis-Cummings Hamiltonian:\cite{Jaynes_Cummings_1963,Tavis_PhysRev_1969}
\begin{widetext}
\begin{equation}
\hat{H}^\text{TC} = \sum^{N_\text{mol}}_j\Delta_j\hat{\sigma}_j^+\hat{\sigma}_j^-+\omega_\text{c}\left(\frac{1}{2}+\hat{a}^\dagger\hat{a}\right)+\sum_j^{N_\text{mol}} g{\boldsymbol{\mu}}_{01}(R_j)\cdot{\boldsymbol{\epsilon}}_\text{c}(\hat{\sigma}^-_j\hat{a}^\dagger+\hat{\sigma}^+_j\hat{a})
\end{equation}
\end{widetext}
Here, $\hat{\sigma}^+_j=\vert1_j\rangle\langle0_j\vert$ and $\hat{\sigma}^-_j=\vert0_j\rangle\langle1_j\vert$ are operators that excite molecule $j$ from the electronic ground state ($\vert0_j\rangle$) into the electronic excited state ($\vert1_j\rangle$) and \textit{vice versa}; $\Delta_j = V_1(R_j) - V_0(R_j)$ is the vertical excitation energy; ${\boldsymbol{\mu}}_{01}(R_j)$ the transition dipole moment of molecule $j$; and ${\boldsymbol{\epsilon}}$ the polarization vector of the cavity vacuum field. As before, the strength of the latter is  $g=\sqrt{\frac{\hbar\omega_c}{2V\epsilon_0}}$.

Within the single excitation manifold, valid for the weak driving typically employed in experiments, this Hamiltonian can be represented in the basis of product states formed from the adiabatic electronic excitations of the molecules and the cavity mode:\cite{luk2017multiscale}
\begin{eqnarray}  \label{eq:arrow_matrix}
    \mathcal{H} = \sum_{j=1}^{N_\text{mol}} V_0(R_j){\bf{1_N}} + 
     \begin{pmatrix}
        \hbar\omega_c  &  \gamma(R_1)  &  \gamma(R_2) & \cdots \\
        \gamma(R_1)    &   \Delta(R_1) &    0         & \cdots \\
        \gamma(R_2)    &    0          &  \Delta(R_2) &  \cdots \\
         \vdots        &   \vdots      &  \vdots      &  \ddots
    \end{pmatrix}  
\end{eqnarray}
where {\bf{1}} is a unit matrix of dimension $N_\text{mol}$, $V_0(R_j)$ is the ground state potential energy of molecule $j$ that is coupled to the cavity through its transition dipole moment: $\gamma(R_j) = g\boldsymbol{\mu}_{01}(R_j)\cdot{\boldsymbol{\epsilon}}_c$.  Intermolecular Coulomb interactions are neglected to isolate cavity-mediated collective effects. 

Diagonalizing the matrix representation of the Tavis-Cummings Hamiltonian (Equation~\ref{eq:arrow_matrix}) in the basis of the single-excitation product states,\cite{Sokolovskii2024b} yields the $N_\text{mol}+1$ adiabatic hybrid light-matter eigenstates:
\begin{equation}
\vert\psi^{m}\rangle=\left(\sum_{j}^{N_\text{mol}}\beta^m_j\hat{\sigma}^+_j + \alpha^m\hat{a}^\dagger\right)\vert\phi_0\rangle
\label{eq:Npolariton}
\end{equation}
with eigenenergies $E_m$. The $\beta_j^m$ and $\alpha^m$ expansion coefficients reflect the contribution of the molecular excitations, $\vert1_j\rangle$, and the cavity mode excitation, $\vert 1_c\rangle$, to eigenstate $\vert \psi^m\rangle$. Due to their parametric dependence on the nuclear degrees of freedom, the eigenenergies form adiabatic potential energy surfaces $E_m({\bf{R}})$, with ${\bf{R}}$ the $3\times N_\text{mol}\times N_\text{atoms}$ coordinates of all atoms in the system.\cite{galego2015cavity,Galego2016PES} Polaritonic surfaces for a single CO molecule coupled to the cavity are shown in Fig.~\ref{fig:CO_pPESs}d. Nonadiabatic couplings between these polaritonic surfaces arise from their nuclear coordinate dependence and are included explicitly in semi-classical treatments.

In such semi-classical treatments, nuclear trajectories are evolved by numerically integrating Newton's equations of motion under the influence of the forces due to the quantum degrees of freedom, for which the wave function, $|\Psi(t)\rangle$, is propagated along the classical trajectory.\cite{Crespo2018} To model the non-adiabatic dynamics in the manifold of eigenstates (Equation~\ref{eq:Npolariton}), we used three popular approaches: (\textit{i}) Ehrenfest; (\textit{ii}) fewest-switches surface hopping\cite{Tully1990,Tully1991} with and (\textit{iii}) without decoherence correction.\cite{Granucci2007critical} The details of their implementation for semi-classical molecular dynamics in the collective strong coupling regime are described in Sokolovskii and Groenhof.\cite{Sokolovskii2024a}

All semi-classical simulations were performed with GROMACS version 4.5.4, using a timestep of 0.1~fs. For each semi-classical method and coupling strength, $g$, 200 non-adiabatic trajectories were propagated for 200~fs.  In the FSSH simulations with decoherence correction, the decoherence parameter, $C$, was 0.1 Hartree (Eq.~17 in Granucci and Persico\cite{Granucci2007critical}). The starting coordinates and velocities were randomly sampled from a Wigner distribution at 0~K and observables were averaged over the trajectories. The 0~K Wigner sampling ensures consistency with the vibrational ground state used in the quantum dynamics simulations.

\section{Results and Discussion}\label{section:results}

\subsection{Bare CO dynamics}

\begin{figure}[!htb]
    \centering
    \includegraphics[width=0.475\textwidth ]{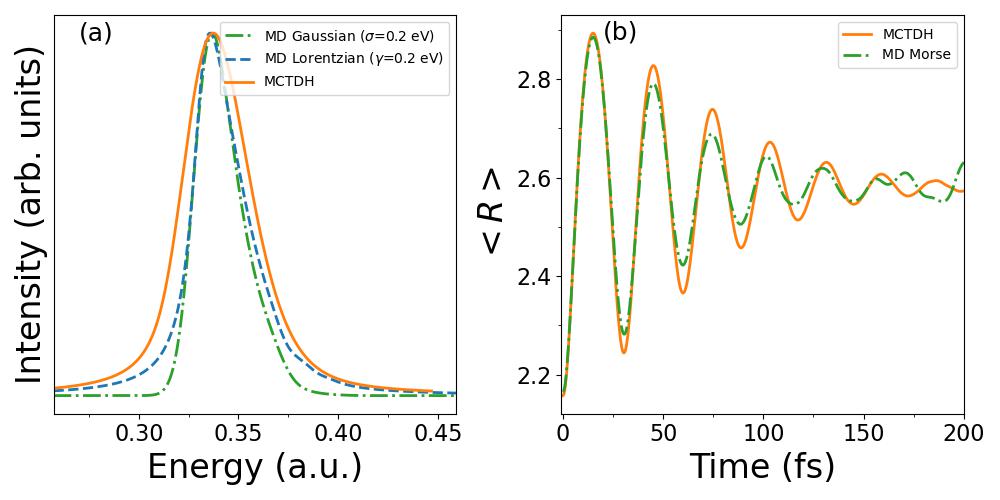}
    \caption{ (a) The linear absorption spectrum of a bare CO molecule of the $^1\Pi$ electronic state is evaluated using both MCTDH (solid line) and MD with Gaussian (dashed dot) and Lorentzian (dashed line) convolution functions. (b) Expectation value of the inter-nuclear distance, $\langle{R}\rangle$, as a function of time after instantaneous excitation into the first excited electronic state of the CO molecule.}
    \label{fig:qexpt-spec}
\end{figure}

Before focusing on the dynamics of CO molecules in a cavity, we first simulate and compare the dynamics of a single CO molecule in vacuum. In Figure~\ref{fig:qexpt-spec}a, we show the linear absorption spectrum of a bare CO molecule obtained with MCTDH and classical molecular dynamics simulations. In the MCTDH simulations, the ground-state vibrational wavefunction, $\vert\psi(0)\rangle$ was promoted from the electronic ground to excited state and propagated for 200~fs. The overlap, $S(t) = \langle\psi(0)\vert\psi(t)\rangle$ was computed, multiplied with an empirical damping function $e^{-t/\tau}$ ($\tau= $~3~fs) and Fourier transformed into the linear absorption spectrum shown in Figure~\ref{fig:qexpt-spec}a. The classical absorption spectrum was obtained as a sum of the vertical excitation energies along MD trajectories, convolved with either a gaussian (width, $\sigma = 0.2$~eV) or a Lorentzian (half-width at half maximum, $\gamma=0.2$~eV) function. Whereas both convolution schemes provide satisfactory agreement with the MCTDH spectrum, the  absorption lineshape matches the MCTDH result more closely with the Gaussian convolution.

In Figure~\ref{fig:qexpt-spec}b, we plot the CO bond length as a function of time after instantaneous excitation from the $^1\Sigma$ electronic ground state into the $^1\Pi$ excited state for both MCTDH and classical MD simulations. The classical result was averaged over 200~trajectories, initiated with different initial positions and velocities, sampled from the Wigner distribution at 0~K in the $^1\Sigma$ electronic ground state. Despite small deviations, which are most apparent at longer timescales, the classical ensemble-averaged $\langle R(t)\rangle$ closely tracks the MCTDH result in the first few vibrational periods. This agreement indicates that for the present 1D model, the classical nuclear approximation is reasonable in vacuum, providing a controlled baseline for assessing nuclear quantum effects under electronic strong coupling.

\subsection{Single molecule strong coupling}

\begin{figure*}[!t]
\centering
\includegraphics[width=\textwidth]{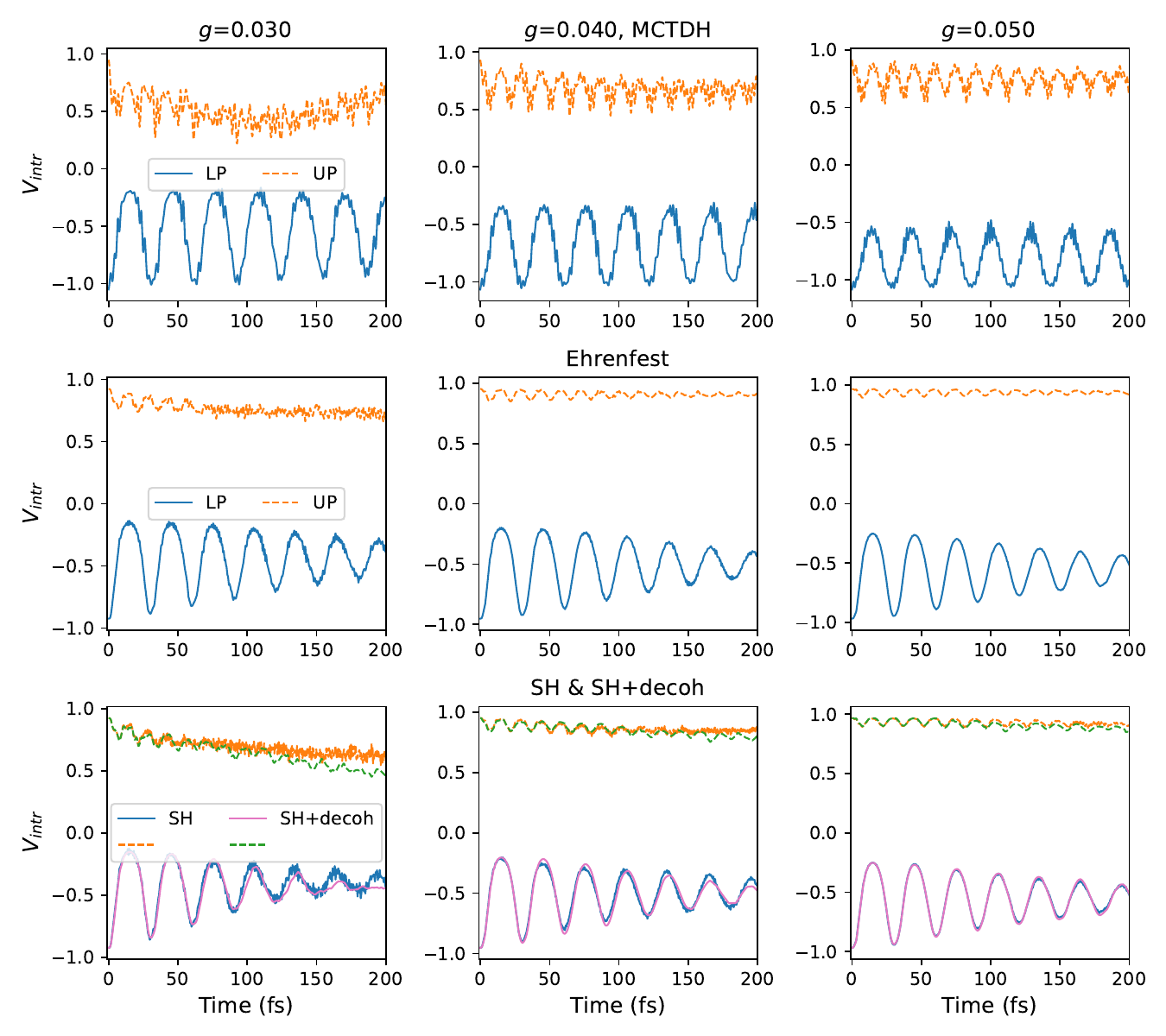}
\caption{\label{fig_vintr_iwp_LP-UP_1CO}The cavity-molecule interaction potential $V_\text{int,r}$, when a single CO molecule is coupled to the cavity mode after vertical excitation into the LP (solid lines) or UP (dashed lines) at different coupling strengths. The MD simulations are averaged over 200 initial conditions sampled from the Wigner distribution of the Morse potential. A value of $V_\text{int,r}$ = -1, or 1 denotes that the population remains trapped in the LP or UP state, respectively.}
\end{figure*} 

When placing a single CO molecule at the anti-node of a cavity mode with a sufficiently strong vacuum field, the molecular and cavity mode excitations hybridize into an upper and lower polariton state, separated by Rabi splitting. Because the Rabi splitting depends critically on the detuning and transition dipole moment, the potential energy surfaces of these hybrid states are different from the uncoupled states.\cite{galego2015cavity} As shown in Fig. \ref{fig:CO_pPESs}d, for a cavity mode on resonance with the vertical excitation energy at the $V_0$ minimum of the bare CO molecule, the surface associated with the LP is lower in energy than the uncoupled $^1\Pi$ excited state ($V_1$), and  has a deeper local minimum, whereas the UP is higher in energy than the uncoupled $^1\Sigma$ ground state ($V_0+\omega_c$). 

To verify if semi-classical MD simulations can capture the effects of these modifications on the nuclear dynamics, we compute the expectation value of the cavity-molecule interaction potential, $V_\text{int,r}$ [cf. Eq. \ref{eq_vint}], and compare to the MCTDH calculation. In Fig. \ref{fig_vintr_iwp_LP-UP_1CO} we show the time-evolution of $V_\text{int,r}$ for a single CO molecule strongly coupled to a cavity mode, after excitation into the LP (solid line) and UP (dashed lines) states at different coupling strengths obtained from full-quantum (MCTDH) simulations (upper row), Ehrenfest MD (middle row) and FSSH MD (bottom row). For all coupling strengths, 
$V_\text{int,r}(t)$ oscillates, reflecting the modulation of the light-matter admixture along the nuclear trajectory as the detuning and transition dipole moment change with bond length. In the LP, the oscillation period and amplitude decrease with increasing coupling strength, consistent with a deeper LP minimum. Both Ehrenfest and FSSH reproduce these qualitative trends. Moreover, $V_\text{int,r}$ remains close to its initial sign following excitation into UP, indicating negligible population transfer into the LP on the simulated timescales. Overall, these results demonstrate that both Ehrenfest and SH methods qualitatively capture the essential features of the polaritonic dynamics in the single molecule strong coupling regime.

\subsection{Multiple molecules}

When more than one molecule interacts with the cavity mode, the collective light-matter coupling modifies the polaritonic states in a nontrivial manner. To ensure that the overall Rabi splitting remains constant when we increase the number of molecules $N_\text{mol}$, we scale down the single-molecule  cavity-molecule coupling constant, $g$, as $g\rightarrow g/\sqrt{N_\text{mol}}$. This scaling maintains the collective coupling strength, while allowing us to systematically investigate cooperative effects arising from multiple molecules inside the cavity. 

First, we consider three \emph{identical} CO molecules strongly coupled to a single cavity mode. Because there is no disorder in the excitation energies, diagonalization of the Tavis-Cummings Hamiltonian $\mathcal{H}$ [cf. Eq. \ref{eq:arrow_matrix}] yields the bright upper (UP) and lower polariton (LP) states as well as two degenerate dark states (DS) at the same energy as the molecular excited states [cf. Fig. S1 of the supplementary information]. In Fig. \ref{fig:3mol_0offset} we plot the time-evolution of the light-matter interaction potential, $V_{\text{int,r}}$, for three collective coupling strengths ($g$ = 0.030, 0.040, and 0.050) in (a-c) MCTDH simulations, (d-f) Ehrenfest MD simulations and (g-i) in FSSH simulations with and without decoherence correction.\cite{Granucci2007critical} To facilitate direct comparison, we start these simulations with a vertical excitation into the upper or lower polaritons.

\begin{figure*}[!htb]
    \centering \includegraphics[width=\textwidth]{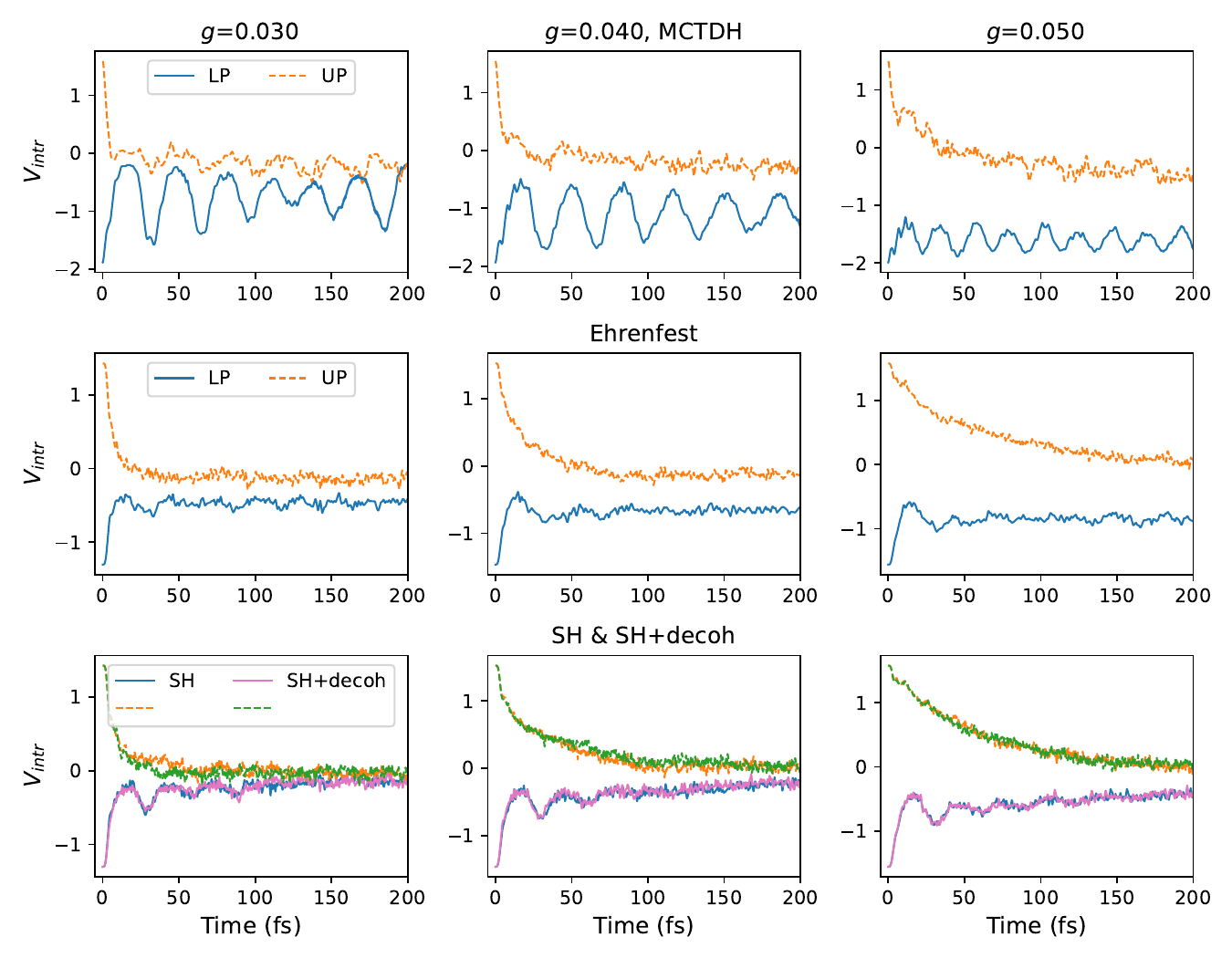}
    \caption{The cavity-molecule interaction potential $V_\text{int,r}$, when the three CO molecules are interacting with the cavity mode after vertical excitation into the LP (solid lines) or  UP (dashed lines) at different coupling strengths.}
    \label{fig:3mol_0offset}
\end{figure*}

In the MCTDH simulations, a vertical excitation into the LP results in an oscillation of $V_\text{int,r}$, suggesting adiabatic evolution between regions with varying photonic contributions to the potential energy surface. As for the single-molecule strong coupling (Fig.~\ref{fig_vintr_iwp_LP-UP_1CO}), the oscillation period and amplitude decrease with increasing coupling strength, reflecting the deepening of the collective LP minimum and the concomitant trapping into this minimum that has been proposed to suppress photochemistry in the strong coupling regime.\cite{Galego2016PES}

In contrast to the single-molecule simulations, an excitation into the UP state rapidly relaxes into the dark state manifold, as indicated by the decay of $V_\text{int,r}$ to zero. The rate of this relaxation decreases with increasing collective coupling strength, or Rabi splitting, reflecting the inverse dependence of the non-adiabatic coupling on the energy gap between the UP and dark state manifold.\cite{Tichauer2022}

In the semi-classical Ehrenfest and FSSH simulations, the relaxation from the UP into the DS manifold is captured qualitatively, indicating that the dominant nonadiabatic transfer pathway is reproduced. In contrast, following excitation into the LP, the oscillations of $V_{\text{int,r}}$ are much more strongly damped in the semi-classical trajectories than in the corresponding MCTDH simulation. The weak sensitivity of this damping to the decoherence correction in the FSSH simulations suggests that it does not primarily originate from the usual over-coherence problem of surface hopping. Instead, it reflects ensemble-induced dephasing: averaging over Wigner-sampled initial nuclear configurations breaks the perfect symmetry of the identical-molecule model and endows the nominally dark states with a small photonic admixture ("gray" states).\cite{Mony2021,Dutta2024,Yin2025} Consequently, weak LP$\rightarrow$DS population transfer becomes allowed in the trajectory ensemble, particularly near geometries where the LP-DS energy gap is minimal.

To enable a more consistent comparison between the fully quantum and trajectory-based descriptions, we next include static energetic disorder. In realistic molecular ensembles disorder is unavoidable, and it also provides a stringent test because it breaks permutation symmetry and mixes bright and dark subspaces. We implement static disorder by shifting the excitation energy of one molecule up by 1\% or 2\% and that of another molecule down by the same amount, keeping the ensemble-average excitation energy unchanged.

\begin{figure*}[!hbt]
    \centering
    \includegraphics[width=\textwidth]{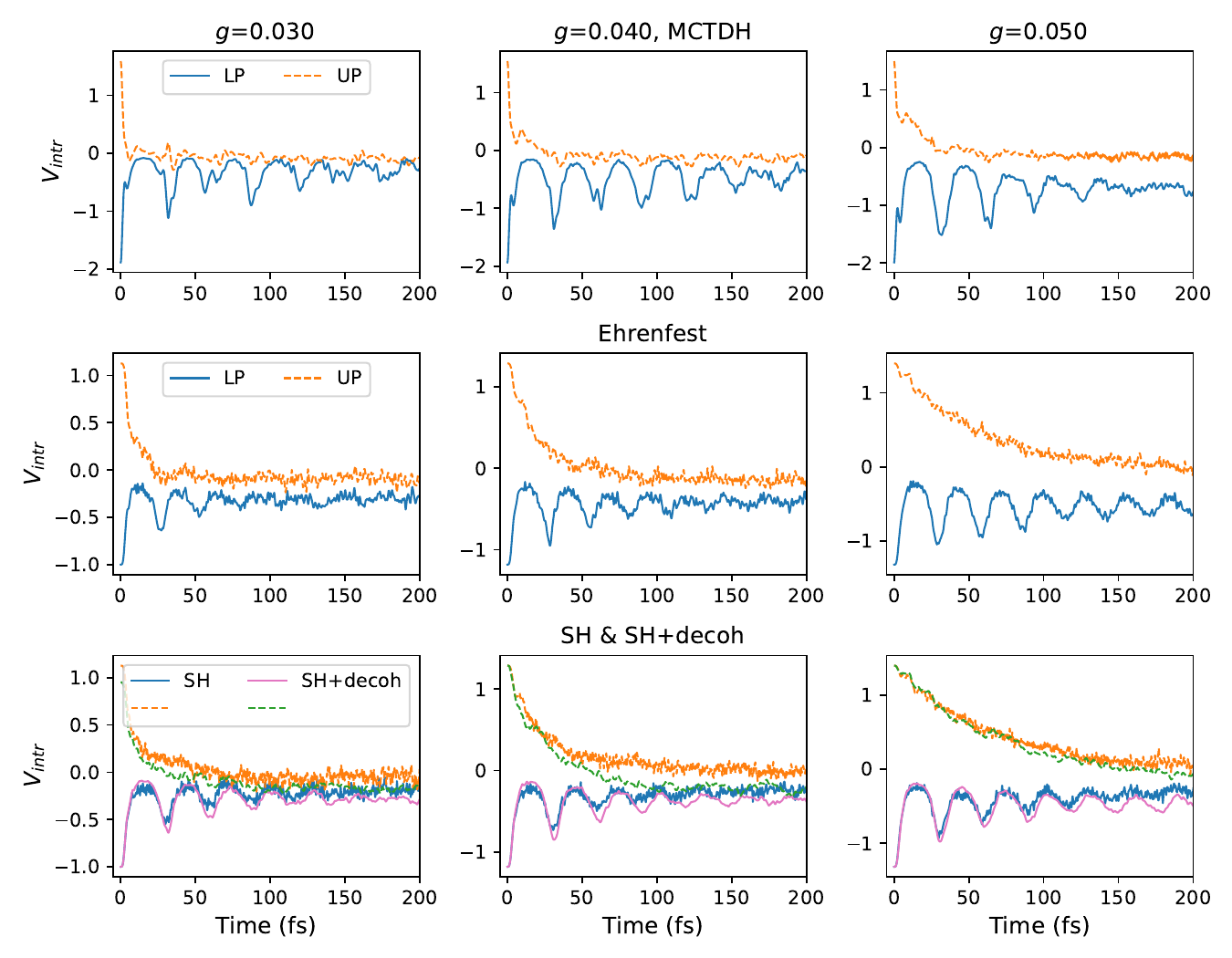}
    \caption{The cavity-molecule interaction potential $V_\text{int,r}$, when the three CO molecules are interacting with the cavity mode after vertical excitation into the LP (solid lines) or UP (dashed lines) states at different coupling strengths. Static disorder ($\pm0.020~E_{\text{S}_1}$)is included such that the excitation energies of the CO molecules are all different.}
    \label{fig:3mol_0o02offset}
\end{figure*}

In Fig.~\ref{fig:3mol_0o02offset}, we plot $V_
\text{int,r}$ for three CO molecules with a 2\% disorder in their excitation energies. In contrast to the situation without disorder (Fig.~\ref{fig:3mol_0offset}), the population injected into the LP upon vertical excitation, dephases in MCTDH simulations, as evidenced by the damping of $V_\text{int,r}$ towards zero. The semi-classical simulations capture this dephasing better than in simulations without static disorder. 

In contrast to the dynamics in the LP,  including disorder has no major effect on the dynamics in the UP. Both with and without disorder, population rapidly transfers into the DS, with a rate that decreases with the Rabi splitting. This relaxation process is also seen in the semi-classical MD simulations, albeit with a lower rate at the higher Rabi splittings. The effect of the decoherence correction is again minor, and only slightly accelerates the population transfer from the UP into the DS at moderate collective coupling strengths. The good qualitative agreement between the full quantum dynamics and semi-classical dynamics simulations for the model with static disorder, is promising for investigating strongly coupled systems that are too complex or large for an MCTDH treatment.

\begin{figure*}[!htb]
    \centering
    \includegraphics[width=1\textwidth ]{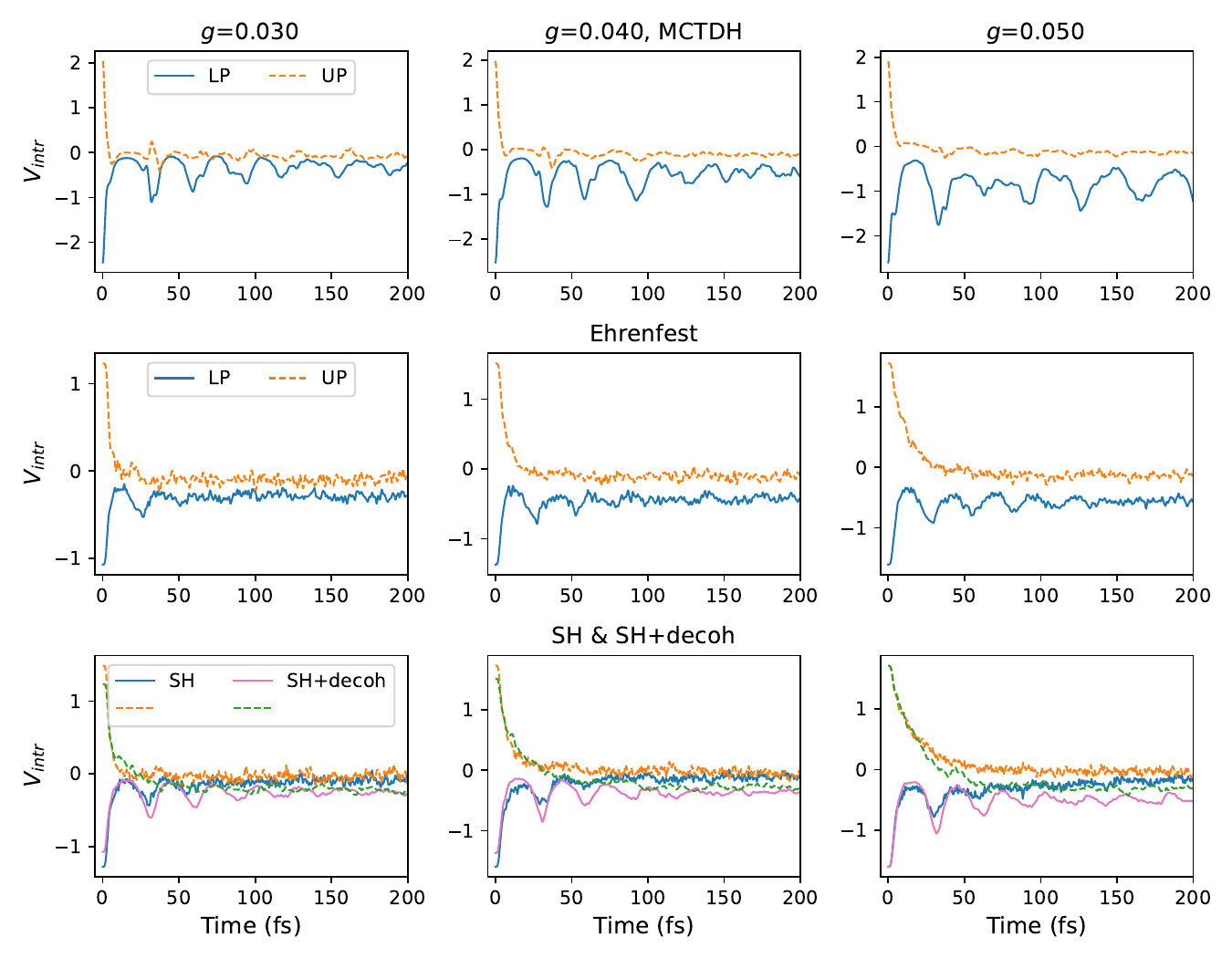}
    \caption{The cavity-molecule interaction potential $V_\text{intr}$, when the five CO molecules are interacting with the cavity mode after vertical excitation into the LP (solid lines) or UP (dashed lines) states at different coupling strengths. Static disorder ($\pm0.020~E_{\text{S}_1}$) is included, such that the excited state energies of the CO molecules are all different.}
    \label{fig:5mol_0o02offset}
\end{figure*}

Finally, we add two more CO molecules and repeat the simulations. As before, we account for static disorder by changing the excitation energies of the five molecules by up to 2\%. The plots of $V_
\text{int,r}$ in Fig.~\ref{fig:5mol_0o02offset}, again reveal decent overall agreement between semi-classical and full quantum dynamics simulations. For 
$N_\text{mol}=5$ with 2\% excitation energy disorder, FSSH yields systematically closer agreement with MCTDH than Ehrenfest dynamics, particularly when the decoherence correction is applied. This trend supports the use of decoherence-corrected FSSH as a practical alternative to fully quantum dynamics for larger electronically strongly coupled ensembles.

\section{Summary and outlook} \label{sec:conclusions}

In this work, we benchmarked mixed quantum-classical molecular dynamics against fully quantum simulations for describing nonadiabatic dynamics under collective electronic strong coupling. Using a model system with up to five CO molecules coupled to a single cavity mode, we directly compared Ehrenfest and Fewest-Switches Surface Hopping (FSSH) dynamics to numerically exact MCTDH results.

For identical molecules without disorder, both semi-classical approaches reproduce the qualitative features of the polaritonic dynamics, including the coherent oscillations following excitation into the lower polariton (LP) and the relaxation from the upper polariton (UP) into the dark-state manifold. When static energetic disorder is included, thereby breaking permutation symmetry and inducing ensemble dephasing, the agreement between semi-classical and fully quantum simulations improves further. Overall, nuclear quantum effects appear to play a minor role for the present CO model in the strong coupling regimes considered here. The largest deviations are observed for the UP$\rightarrow$dark-state relaxation, which proceeds slightly faster in the fully quantum treatment and remains discernible as the number of molecules increases.

Between the two mixed quantum-classical approaches, FSSH provides systematically closer agreement with MCTDH than Ehrenfest dynamics, particularly when a decoherence correction is applied.\cite{Granucci2007critical} This finding supports the use of decoherence-corrected FSSH as a practical and computationally efficient alternative for simulating electronically strongly coupled molecular ensembles beyond the reach of exact quantum dynamics.

Finally, we assessed the influence of the specific light-matter Hamiltonian employed. In the fully quantum MCTDH simulations, the light–matter interaction is described using the Rabi Hamiltonian, which includes counter-rotating terms. In contrast, the semi-classical simulations employ the Tavis-Cummings Hamiltonian within the rotating-wave approximation. For the moderate coupling strengths considered here ($g<<\omega_c$), counter-rotating contributions lead only to small corrections to the polariton energies (Bloch-Siegert shifts) and do not qualitatively affect the nonadiabatic dynamics. The good agreement observed between the two approaches supports this expectation. In the ultra-strong or deep-strong coupling regimes, these contributions are expected to become more important;\cite{DeLiberato2014,Forn-Diaz2019,DeBernardis24} however, such regimes are considerably more challenging to access experimentally for electronic transitions.\cite{Frisk2019,Meuller2020}

Taken together, our results demonstrate that mixed quantum-classical methods, and in particular decoherence-corrected FSSH, provide a reliable and scalable framework for modeling nonadiabatic photochemistry under collective electronic strong coupling in systems that are too large for fully quantum mechanical treatments.

\begin{acknowledgments}
We thank Ilia Sokolovskii for 
valuable discussions and for his help in setting up the MD simulations. This work was supported by Finnish Cultural Foundation (Grant No. 00231164). The CSC-IT center for scientific computing in Espoo, Finland, is acknowledged for providing very generous computational resources. 

\end{acknowledgments}

\section*{Data Availability Statement}
The data that support the findings of this study are available from the corresponding author upon reasonable request. The code, based on a fork of Gromacs-4.5.3, is available for download at https://github.com/upper-polariton/GMXTC.git

\section*{Supporting Information}
The Supporting Information contains a plot of the potential energy profiles of three identical CO molecules coupled to a cavity mode.

\bibliography{main}

\end{document}


\centering
\title{Supporting information \\ for \\ Benchmarking mixed quantum-classical dynamics for collective electronic strong coupling }
\author{Arun Kumar Kanakati$^1$, Oriol Vendrell$^2$ and Gerrit Groenhof$^1$}
\thanks{Electronic mail: arun.k.kanakati@jyu.fi ; gerrit.x.groenhof@jyu.fi  }
\affiliation{$^1$Department of Chemistry and Nanoscience Center, University of Jyv\"{a}skyl\"{a}, 40014 Jyv\"{a}skyl\"{a}, Finland. \\
$^2$Theoretical Chemistry, Institute of Physical Chemistry, Heidelberg University, Im Neuenheimer Feld 229, 69120 Heidelberg,
Germany}

\maketitle
\renewcommand\thetable{S\arabic{table}}
\renewcommand\thefigure{S\arabic{figure}}

\renewcommand\thefigure{S1}
\begin{figure}[h!]
\centering
\includegraphics[scale=0.6, angle=0]{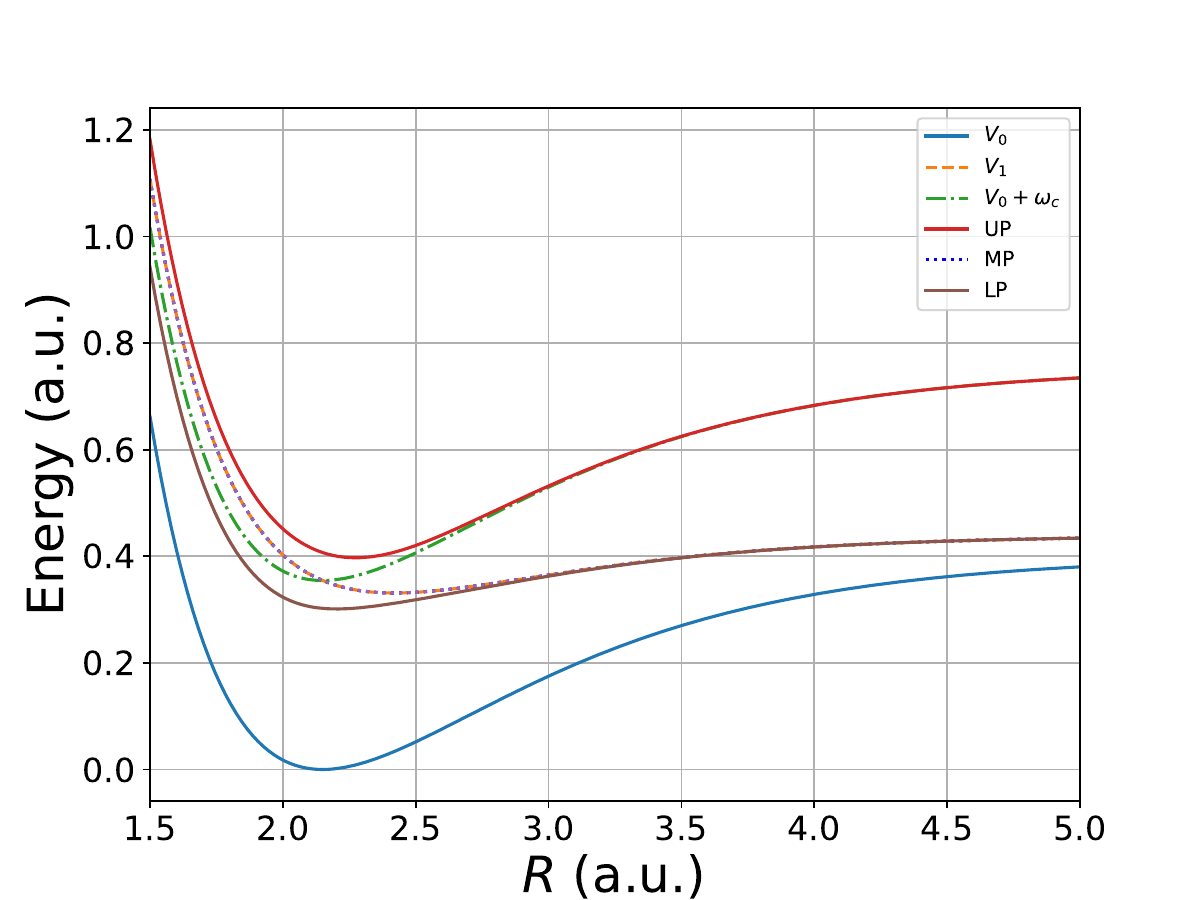}
\caption{\label{fig_pPESs_3CO}Potential energy profiles of three CO molecules in a single cavity mode. The coupling strength considered here is 0.050 au. The Profiles without coupling show the energy of the ground state shifted to the energy mode of the cavity (i.e. cavity mode excitation, $V_0+\omega_c$), as well as the ground and excited states of the CO molecule. The hybrid states as result of the coupling between the cavity and the molecule. These cuts are obtained by the diagonalization of the cavity and electronic Hamiltonian [cf. Eq. 13]. }
\end{figure} 